\documentclass[%
 reprint,
 amsmath,amssymb,
 aps,
 fleqn
]{revtex4-1}

\usepackage{graphicx}% Include figure files
\usepackage{dcolumn}% Align table columns on decimal point
\usepackage{bm}% bold math
\begin{document}

\preprint{APS/123-QED}

\title{Observation of Mollow triplet with metastability exchange collisions in $^{3}\textrm{He}$ atoms}% Force line breaks with \\

\author{Yuanzhi Zhan}
\author{Xiang Peng}%
\email{xiangpeng@pku.edu.cn}
\author{Sheng Li}
\author{Liang Zhang}
\author{Jingbiao Chen}
 \email{jbchen@pku.edu.cn}
%%% \altaffiliation[Also at ]{Physics Department, XYZ University.}%Lines break automatically or can be forced with \\
\author{Hong Guo}%
 \email{hongguo@pku.edu.cn}
\affiliation{%
State Key Laboratory of Advanced Optical Communication Systems and Networks, School of Electronics Engineering and Computer Science, and Center for Quantum Information Technology, Peking University, Beijing 100871, China\\
\\
}%


\date{\today}% It is always \today, today,

\begin{abstract}
We study the dressed states of $^{3}\textrm{He}$ atoms and experimentally observe the Mollow triplet (MT) induced with an ultra-low-frequency (ULF) oscillating magnetic field as low as 4~Hz. The ULF-MT signatures from the ground states of $^{3}\textrm{He}$ atoms are transferred to the metastable states by metastability-exchange collisions (MECs) and measured optically, which demonstrates 2~s coherence time in the dressed ground states. The result shows the possibility of ULF magnetic field amplitude measurement and a new scheme for optical frequency modulation.
%\begin{description}
%42.50.Gy, 32.80.Qk, 32.30.−r
%\end{description}
\end{abstract}

%\pacs{78.67.Hc, 42.50.Ct, 42.50.Hz, 78.55.-m}% PACS, the Physics and Astronomy
\maketitle

%\tableofcontents

\section{\label{sec1}INTRODUCTION}
Mollow triplet (MT) is a kind of coherence between dressed states in two-level systems, which provides a method for detecting the internal dynamics in trap \cite{leibfried2003quantum}, and leads to the development of applications in laser cooling \cite{kim2013mollow} and quantum information processing \cite{peiris2017franson}. Mollow triplet was originally introduced as a phenomenon that a monochromatic optical field coupling a two-level atomic system induces three peaks of scattered fluorescence by the electric-dipole (ED) interactions \cite{mollow1969power}. The MT of resonant light scattering was further investigated in the atomic beam of sodium \cite{wu1975investigation}, quantum dot \cite{fischer2016self}, silicon vacancy of diamond \cite{zhou2017coherent}, and superconducting circuits \cite{baur2009measurement}.

\begin{figure}
	\includegraphics[width=0.4\textwidth]{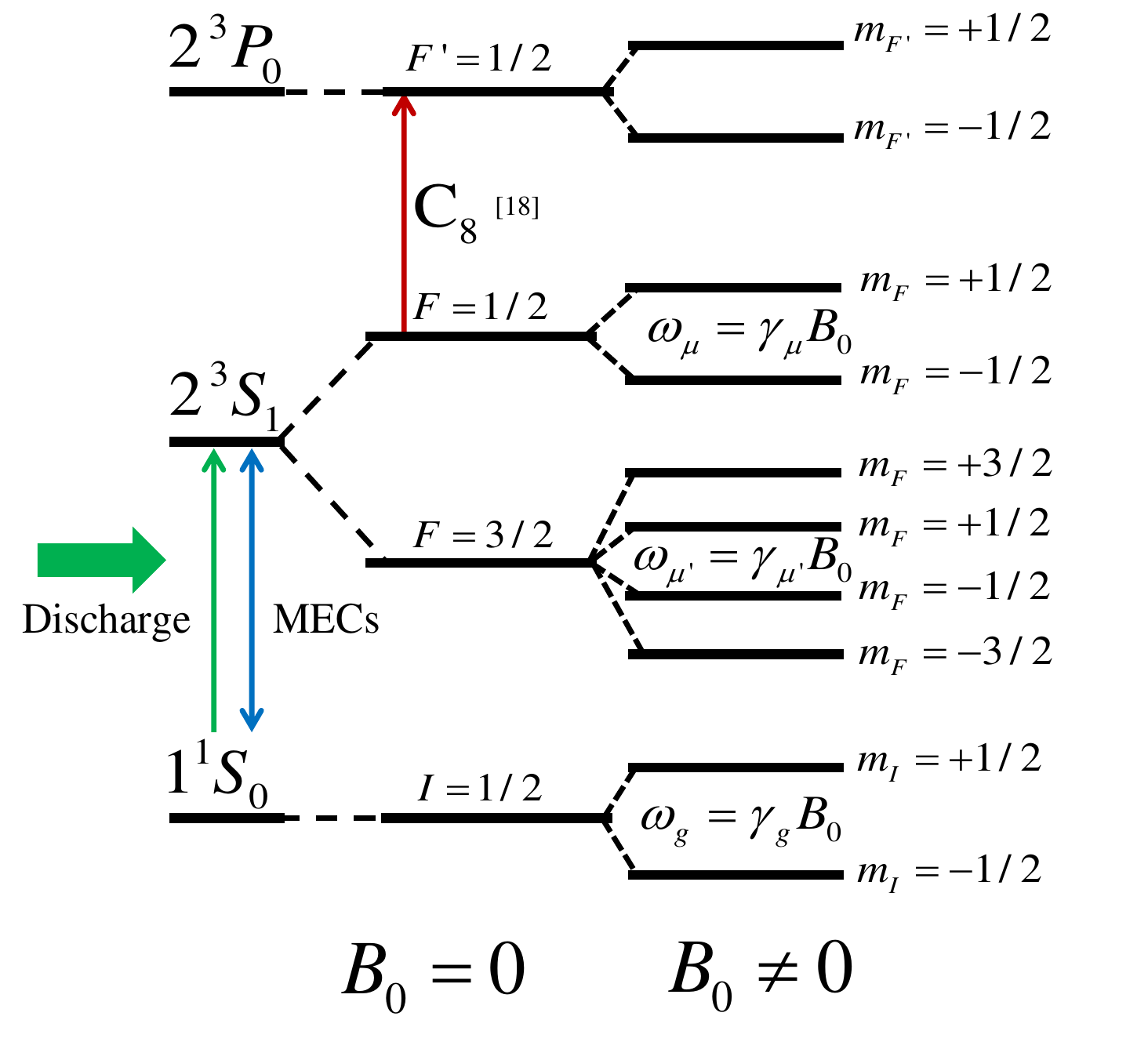}% Here is how to import EPS art
	\caption{\label{fig:energy} The energy-level diagram of $^{3}\textrm{He}$ atoms (not in scale). $1^{1}S_{0}$, $2^{3}S_{1}$, $2^{3}P_{0}$ are the ground state, metastable state, excited state, respectively. The metastable state contains two hyperfine states, $F=1/2$ and $F=3/2$, which are split into 2 and 4 Zeeman sublevels in a static magnetic field $\bf{B_0}$ respectively. The blue line indicates the metastability exchange collisions (MECs) between $1^{1}S_{0}$ and $2^{3}S_{1}$. The interval depends on the gyromagnetic ratio, i.e. $\gamma_{\rm g}=$~3.2~kHz/G for ground state, $\gamma_\mu=$~3.8~MHz/G and $\gamma_{\mu'}=$~1.9~MHz/G for $F=1/2$, $F=3/2$, respectively \cite{colegrove1963polarization}. The green line indicates the RF discharge to generate $2^{3}S_{1}$ atoms. The red line indicates the optical transition $\textrm{C}_8$ between the $2^{3}S_{1}$ and $2^{3}P_{0}$ states with the vacuum wavelength 1083.353 nm, which is used for optical pumping and probing \cite{nacher1985optical}.
	}
\end{figure}

The MTs in these previous works were mainly induced by the ED interaction with electromagnetic fields whose frequencies ($f_{\rm ED}$) cover from microwave to optical range. In principle, the microwave radiation is a preferable choice for applications that require a longer coherence time \cite{mintert2001ion}, due to the fact that the coherence time of quantum states is limited by the spontaneous emission rate\cite{ozeri2007errors, PhysRevA.69.042308}. The microwave-induced MT with the magnetic-dipole (MD) interaction has been observed in superconducting loop \cite{astafiev2010resonance}, Nitrogen Vacancy centers \cite{rohr2014synchronizing}, spin-nanomechanical systems \cite{pigeau2015observation}. The radiowave-induced MT, with the coherence time of 5~ms, has been demonstrated in an electromagnetically induced transparency system of $^{87}\textrm{Rb}$ \cite{basler2015radio}.

In this article, we utilize $^{3}\textrm{He}$ atoms to achieve the ultra-low-frequency (ULF) MT induced by the MD interaction. The ground state of $^{3}\textrm{He}$ atoms has the advantages of smaller gyromagnetic ratio and longer coherence time than that of $^{87}\textrm{Rb}$, and thus can achieve the ULF MT. The MT signal can be observed at the condition that Rabi frequency of the oscillating field is far greater than the transverse relaxation rate \cite{mollow1977elastic}. After the spontaneous emission rate decreases low enough, the other relaxation process (like the optical pump, the collisions among atoms, the collisions between atoms and wall, the gradient of magnetic field) will be dominant in the relaxation mechanism. The energy-level diagram of $^{3}\textrm{He}$ atoms is illustrated in Fig.~\ref{fig:energy} \cite{gentile2017optically}. The traditional MT in sodium was observed through laser-induced resonance fluorescence \cite{wu1975investigation}. However this method cannot be used for observing the ground-state MT of $^{3}\textrm{He}$, because the optical transitions are not easily accessible. It is also challenging to detect the ULF MT signal from the ground state directly with a pick-up coil \cite{esler2007dressed,chu2011dressed}. Here we use the metastability exchange collisions (MECs) to transfer the MT of $1^{1}S_{0}$ to the metastable state $2^{3}S_{1}$ generated through a discharge, and then detect it in $2^{3}S_{1}$ with the optical method. Therefore we observe the ULF MT induced with a 4~Hz oscillating magnetic field coupling the ground-state $^{3}\textrm{He}$ atoms, and the coherence or transverse relaxation time of the ULF MT around 2~s at room temperature.

The paper is organized as follows. In Sec.~\ref{sec2} we describe the dressed-states physical picture of the MT transfer by the MECs, and use the angular momentum equations to simulate the process. In Sec.~\ref{sec3} we introduce the ULF MT experiment, including the observed MT signal both in the time and frequency domain. Based on this, we investigate the influence of the oscillating magnetic field and the pumping beam on ULF MT in $^{3}\textrm{He}$ atoms. The simulation results are in a good agreement with the experiments data. The applications of our results are discussed in Sec.~\ref{sec4}.

\section{\label{sec2}MODEL}
\begin{figure}[t]
	\includegraphics[width=0.45\textwidth]{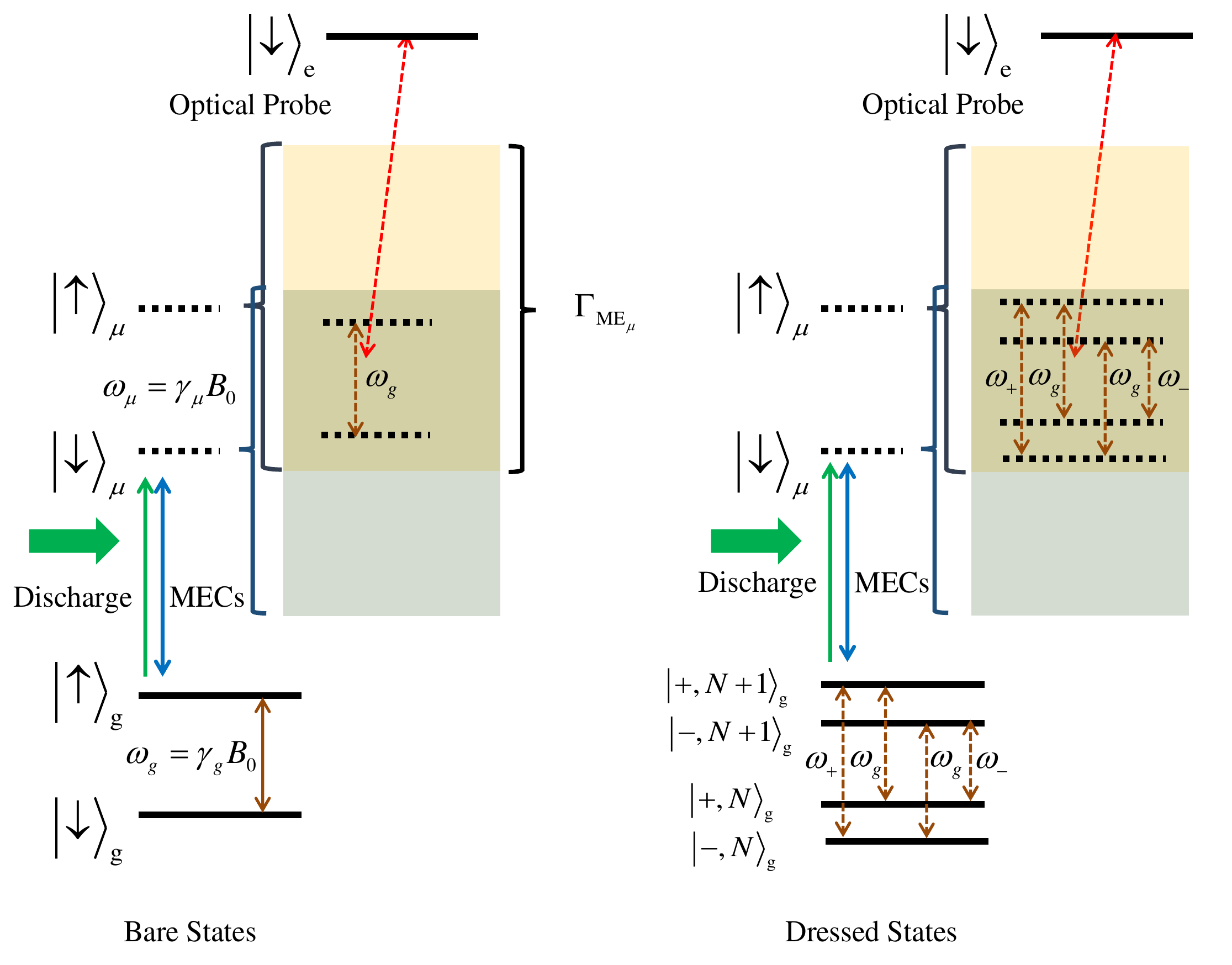}% Here is how to import EPS art
	\caption{\label{fig:dressedspin}The principle of detecting the ultra-low-frequency Mollow triplet (ULF MT) which is transferred via MECs. The subscript g ($\mu$, e) represents the ground state (metastable state, excited state). The dark square means the collision broadening of the energy levels by MECs, which communicate the ground state with the metastable states. The red dashed line represents the coupling of the metastable state and the excited states by optical transition.
	}
\end{figure}

We adopt the dressed atom approach to describe the physical picture \cite{cohenbook}. Figure~\ref{fig:dressedspin} illustrates the MT transfer from the ground state to the metastable state. The oscillating magnetic field drives the ground-state coherence of two Zeeman energy eigenstates or bare states $\vert$$\uparrow$$\rangle _{\rm g}$ and $\vert$$\downarrow$$\rangle _{\rm g}$. The new energy eigenstates or a series of dressed states are formed as the superposition states $\vert$$+$,$N$$\rangle_{\rm g}$$=(1/\sqrt{2})$$($$\vert$$\downarrow$,$n$$\rangle_{\rm g}$$+$$\vert$$\uparrow$,$n-1$$\rangle_{\rm g}$$)$ and $\vert$$-$,$N$$\rangle_{\rm g}$$=(1/\sqrt{2})$$($$\vert$$\downarrow$,$n$$\rangle_{\rm g}$$-$$\vert$$\uparrow$,$n-1$$\rangle_{\rm g}$$)$, where $n$ is the quantum number of oscillating field, $\vert$$\uparrow$,$n$$\rangle_{\rm g}$ ($\vert$$\downarrow$,$n$$\rangle_{\rm g}$) is the direct product state of atoms and the oscillating field and $N$ is the total number of the excitations in the system. Note that the energy interval $\hbar \Omega_{\rm R} = (1/2)\hbar\gamma_{\rm g}B_{\rm M}$, where $B_{\rm M}$ is the amplitude of the oscillating magnetic field. There are three transition frequencies between $\Delta N=\pm 1$ superposition states, i.e., $\omega_{\rm g}$ between $\vert +,N+1 \rangle _{\rm g}$ ($\vert -,N+1 \rangle _{\rm g}$) and $\vert +,N \rangle _{\rm g}$ ($\vert -,N \rangle _{\rm g}$), $\omega _+=\omega _{\rm g} +\Omega _{\rm R}$ between $\vert +,N+1 \rangle _{\rm g}$ and $\vert -,N \rangle _{\rm g}$ and $\omega _-=\omega _{\rm g} -\Omega _{\rm R}$ between $\vert +,N \rangle _{\rm g}$ and $\vert -,N+1 \rangle _{\rm g}$, which are called Mollow spectrum or Mollow triplet \cite{mollow1969power}. At the same time, the oscillating magnetic field cannot drive the metastable states, considering that the gyromagnetic ratio of the metastable state is thousands times larger than that of the ground state. Through the MECs, the coherence can be transferred from the ground state to the metastable states on the condition of $\omega_{\mu} -\omega_{\rm g} <\Gamma_{\textrm{ME}_{\mu}}$ \cite{partridge1966transfer}, where $\omega_{\mu}$ is the Larmor frequency of $F = 1/2$ metastable state and $\Gamma_{\textrm{ME}_{\mu}}$ is the MECs rate between $F = 1/2$ state and the ground state. In our case, with 100 nT magnetic field $B_0$, $\Gamma_{\textrm{ME}_{\mu}}$ is approximately estimated as 1~MHz for the 1~Torr $^{3}\textrm{He}$ atomic cell at room temperature \cite{nacher1985optical}, which is much larger than $\omega_{\mu}-\omega_{g}\approx $ 3.8 kHz. Thus the coherence of the metastable states induced by MECs can be detected through an optical transitions such as $\textrm{C}_8$ shown in Fig.~\ref{fig:energy}.

The ground-state $^{3}\textrm{He}$ atom interacting with the magnetic field can be described by the following Hamiltonian
\begin{equation}
\qquad   \hat{H}_I =\hbar\gamma _{\rm g} B_0 \hat{I}_z   - \hbar\gamma _{\rm g} B_{\rm M} \hat{I}_y \textrm{cos}\omega t ,
\label{eq:ham}
\end{equation}
where $\hat{I}_z$ ($\hat{I}_y$) is the nuclear angular momentum operators  projecting in the direction of $z$ $(y)$ axis or the static (oscillating) magnetic field $B_0$ ($B_{\rm M}$) and $\omega$ is the frequency of the oscillating magnetic field. The dynamics of the ground-state and metastable-state atoms can be described with the Bloch equations following as 

\begin{equation}
\begin{split}
\frac{d}{{dt}}{\bf{I}} = & {\gamma _{\rm g}}({B_{\rm M}}\cos \omega t{{\bf{e}}_y} + {B_0}{{\bf{e}}_z}) \times {\bf{I}}\\
 + & \Gamma_{\textrm{ME}_\textrm{g}}( - {\bf{I}}  - \frac{{\rm{1}}}{{\rm{3}}}{{\bf{F}}_{\mu}}+ \frac{1}{3}{{\bf{F}}_{\mu'}})
  -  {\Gamma _{\textrm{g}}}{\bf{I}}, 
\end{split}
\label{eq:DEq0}
\end{equation}

\begin{equation}
\begin{split}
\frac{d}{{dt}}{{\bf{F}}_{\mu}} = & {\gamma _{{\mu}}}({B_{\rm M}}\cos \omega t{{\bf{e}}_y} + {B_0}{{\bf{e}}_z}) \times {{\bf{F}}_{\rm{\mu}}} \\
+ & \Gamma_{\textrm{ME}_{\mu}}( - \frac{7}{9}{{\bf{F}}_{\mu}} + \frac{{{\rm{1}}}}{9}{{\bf{F}}_{\mu'}} - \frac{{1}}{9}{\bf{I}}) \\
- & {\Gamma _{\mu}}({{\bf{F}}_{\mu}}- {P}{{\bf{e}}_z}), 
\end{split}
\label{eq:DEq1}
\end{equation}

\begin{equation}
\begin{split}
\frac{d}{{dt}}{{\bf{F}}_{\mu'}} = & {\gamma _{{\mu'}}}({B_{\rm M}}\cos \omega t{{\bf{e}}_y} + {B_0}{{\bf{e}}_z}) \times {{\bf{F}}_{\mu'}}\\
 + & \Gamma_{\textrm{ME}_{\mu'}}( - \frac{4}{9}{{\bf{F}}_{\mu'}} + \frac{{\rm{10}}}{9}{{\bf{F}}_{\mu}} + \frac{10}{9}{\bf{I}})\\ 
  - & {\Gamma _{\mu'}}{{\bf{F}}_{\mu'}}, 
\end{split}
\label{eq:DEq2}
\end{equation}
respectively, where the index g, $\mu$, $\mu'$ are corresponding to the ground state, $F=1/2$ and $F=3/2$ metastable states, $\bf{e}_z$ or $\bf{e}_y$ is the unit vector of z or y axis, $\bf{I}=\textrm{Tr}(\rho \hat{I})$, $\bf{F}_{\mu}=\textrm{Tr}(\rho \hat{F}_{\mu})$, $\bf{F}_{\mu'}=\textrm{Tr}(\rho \hat{F}_{\mu'})$ are the angular momentum expectations of ground state and two metastable states, $\rho$ is the density matrix of the atomic system, $\Gamma_{\textrm{ME}_{\rm g}}$, $\Gamma_{\textrm{ME}_{\mu}}$, $\Gamma_{\textrm{ME}_{\mu'}}$ are the MECs rates for the different states. $\Gamma_{\rm g}$, $\Gamma_{\mu}$, $\Gamma_{\mu'}$ are the transverse relaxation rate (decoherence rate) includes the effect of the spontaneous emission, the collisions between atoms and wall, the RF discharge and the magnetic field gradient, but excepts for the MECs and the pump effect. We have assumed the longitudinal relaxation rate is equal to the transverse relaxation rate for simulation. The first terms of these equations describe the interaction between the spin ensemble and the magnetic field. The second terms describe the metastability exchange between the ground state and metastable states. The third terms describe the other relaxation of each states. As the MECs rate $\Gamma_{\textrm{ME}_{\mu(\mu')}}$ (1~MHz) for the metastable state is much larger than the MECs rate $\Gamma_{\textrm{ME}_{\rm g}}$ and the Larmor frequency $\omega_{\rm g}$ of the ground state in $B_0\approx$~100~nT, the motion of metastable-state angular momentum can be treated as the quasi static compared with the motion of ground-state angular momentum. Additionally, as the oscillating magnetic field is far from detuning for the Larmor frequency of the metastable states $\left| \omega-\omega_{\mu(\mu')} \gg 0 \right|$, and $\Gamma_{\textrm{ME}_{\mu(\mu')}}$ is much larger than the relaxation rate $\Gamma_\mu$ (1~kHz), $\gamma_{\mu(\mu')} B_0$ (1~kHz) and $\gamma_{\mu(\mu')}  B_{\rm M}$, the effect of the magnetic field on the metastable state actually can be ignored. Therefore we can obtain the evolution equation only depends on the angular momentum of the ground states

\begin{equation}
\begin{split}
\frac{d}{{dt}}{\bf{I}} = & {\gamma _{\rm g}}({B_{\rm M}}\cos \omega t{{\bf{e}}_y} + {B_0}{{\bf{e}}_z}) \times {\bf{I}}\\
+ & \frac{\Gamma_{{\rm ME}_{\rm g}}\Gamma_{\mu}}{\Gamma_{{\rm ME}_{\mu(\mu')}}}{P}{{\bf{e}}_z}
-  {\Gamma _{\rm g}}{\bf{I}}. 
\end{split}
\end{equation}
Notice that the second term related the MECs rates and the pump effect is a constant, and the relaxation rate of ground-state angular momentum only depends on $\Gamma_{\rm g}$. In other words, $\Gamma_{\rm g}$ is approximately equal to the transverse relaxation rate of the ground states $\Gamma_2$ at the experimental condition. Atomic angular momentum polarization is generated in the metastable state by the pumping light and transferred to the ground state via the MECs, while the MT signal is generated in the ground state and transferred back to the metastable state by the MECs and measured with the optical field. Contributing from the low spontaneous emission rate of low frequency magnetic moment transition, the coherence time of the superposition states $\vert +,N \rangle _{\rm g}$ and $\vert -,N \rangle _{\rm g}$ is determined by other decoherence mechanism like the optical pump, the collisions among $^{3}\textrm{He}$ atoms, the collisions with the wall, and the magnetic field gradients. Equations~(\ref{eq:DEq0}-\ref{eq:DEq2}) can be numerically solved and compared with the experimental results. The related parameters for the simulation are listed in Tab.~\ref{tab:table1}.

\begin{table}[t]
	\caption{\label{tab:table1}%
		The simulation parameters for Equations~(\ref{eq:DEq0}-\ref{eq:DEq1}) list here. The index g, $\mu$, $\mu'$ are corresponding to ground state, $F=1/2$ and $F=3/2$ metastable states. $\gamma_{\rm g}=$~3.2~kHz/G for ground state, $\gamma_\mu=$~3.8~MHz/G, $\gamma_{\mu'}=$~1.9~MHz/G for $F=1/2$, $F=3/2$ are the gyromagnetic ratio \cite{colegrove1963polarization}, $\Gamma_{\textrm{ME}_{\rm g}}=1$, $\Gamma_{\textrm{ME}_{\mu}}=10^6$, $\Gamma_{\textrm{ME}_{\mu'}}=10^6$ are the MECs rates for the different states \cite{nacher1985optical}. $\Gamma_{\rm g}$, $\Gamma_{\mu}$, $\Gamma_{\mu'}$ are the transverse relaxation rate excepts for the MECs and pump effect. The value of $\Gamma_{\rm g} \approx \Gamma_2 =0.5$ Hz is measured through traditional magnetic resonance methods, and the values of $\Gamma_{\mu}$, $\Gamma_{\mu'}$ is limited by the lifetime of the metastable states near 1~ms.}
	\begin{ruledtabular}
		\begin{tabular}{ccc|cccccc}
			$\gamma_g$&$\gamma_{\mu}$&$\gamma_{\mu'}$&			$\Gamma_{\textrm{ME}_{\rm g}}$&$\Gamma_{\textrm{ME}_{\mu}}$&$\Gamma_{\textrm{ME}_{\mu'}}$&			$\Gamma_{\rm g}$&$\Gamma_{\mu}$&$\Gamma_{\mu'}$\\
			~&(MHz/G)&~&~&~&(Hz)&~&~&~\\			
			\hline
			0.0032&3.8&1.9&1&$10^6$&$10^6$&0.5&$10^3$&$10^3$\\		
		\end{tabular}
	\end{ruledtabular}
\end{table}

\section{\label{sec3}EXPERIMENTAL RESULTS}
\begin{figure}[t]
	\includegraphics[width=0.45\textwidth]{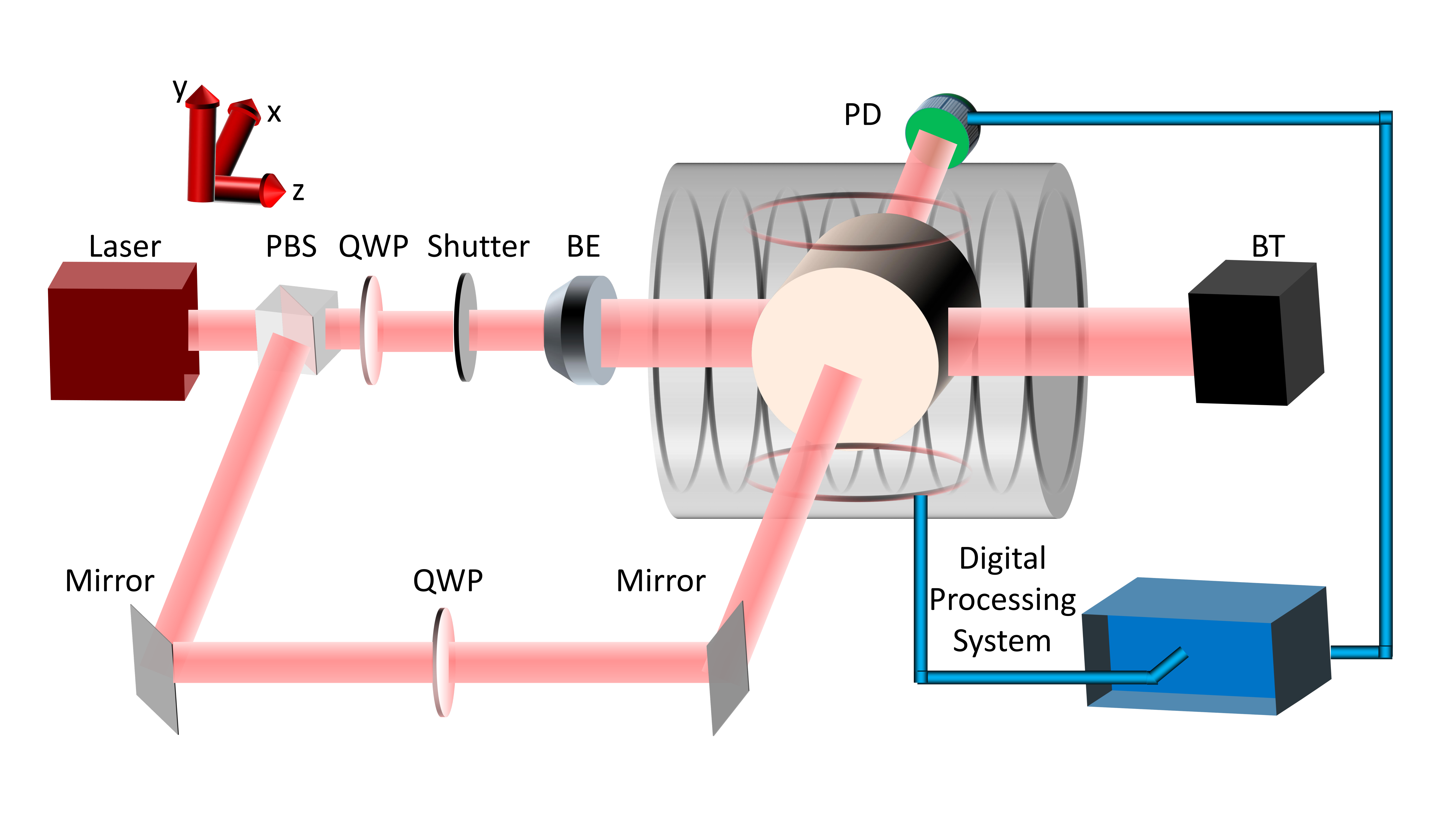}% Here is how to import EPS art
	\caption{\label{fig:experiment} The experimental setup for the ULF-MT measurement in $^{3}\textrm{He}$ atoms. PBS: Polarization Beam Splitter, QWP: Quarter Wave Plate, HWP: Half Wave Plate, BE: Beam Expander, BT: Beam Trap, PD: Photodetector.
	}
\end{figure}

We use the optical method to detect the MT of the metastable state which is coupled with the ground state. The experiment setup is shown in Fig.~\ref{fig:experiment}. Both the pump and the probe beams are from a narrow-linewidth laser source (NKT Photonics Y10 \& LEA Photonics MLXX-EYFA-CW-SLM-P-TKS) of which the center frequency is tuned to the $\textrm{C}_8$ transition of $^{3}\textrm{He}$ atoms. The pump (probe) beam propagates along $z (x)$ axis and has the power of 50~mW (0.3~mW), with $1/e^2$ waist diameter of about 20~mm (1~mm). Both the beams are circularly polarized before entering the cell. The optical detection has higher signal to noise ratio compared with that of pick-up coils for low-frequency MT signal. The home-made pure $^{3}\textrm{He}$ (pressure: 0.6~Torr) cylindrical atomic cell (size: $\phi$50$\times$L70 mm$^3$) is located in the seven-layer magnetic shield, and is excited by a radio-frequency power source (50 MHz, 0.8~W) to continuously discharge and generate the metastable-state atoms. The solenoid generates the static magnetic field $B_0$ along $z$, and a set of helmholtz coil generates the oscillating magnetic field $B_{\rm M}$ along $y$. The digital processing system includes the PXI-4461 and PXI-4462 (resolution: 24-Bit, sampling rate: 204.8 kS/s) of the National Instruments, which is used for controlling the helmholtz coil and signal acquisition.
 
\begin{figure*}[t]
	\includegraphics[width=0.9\textwidth]{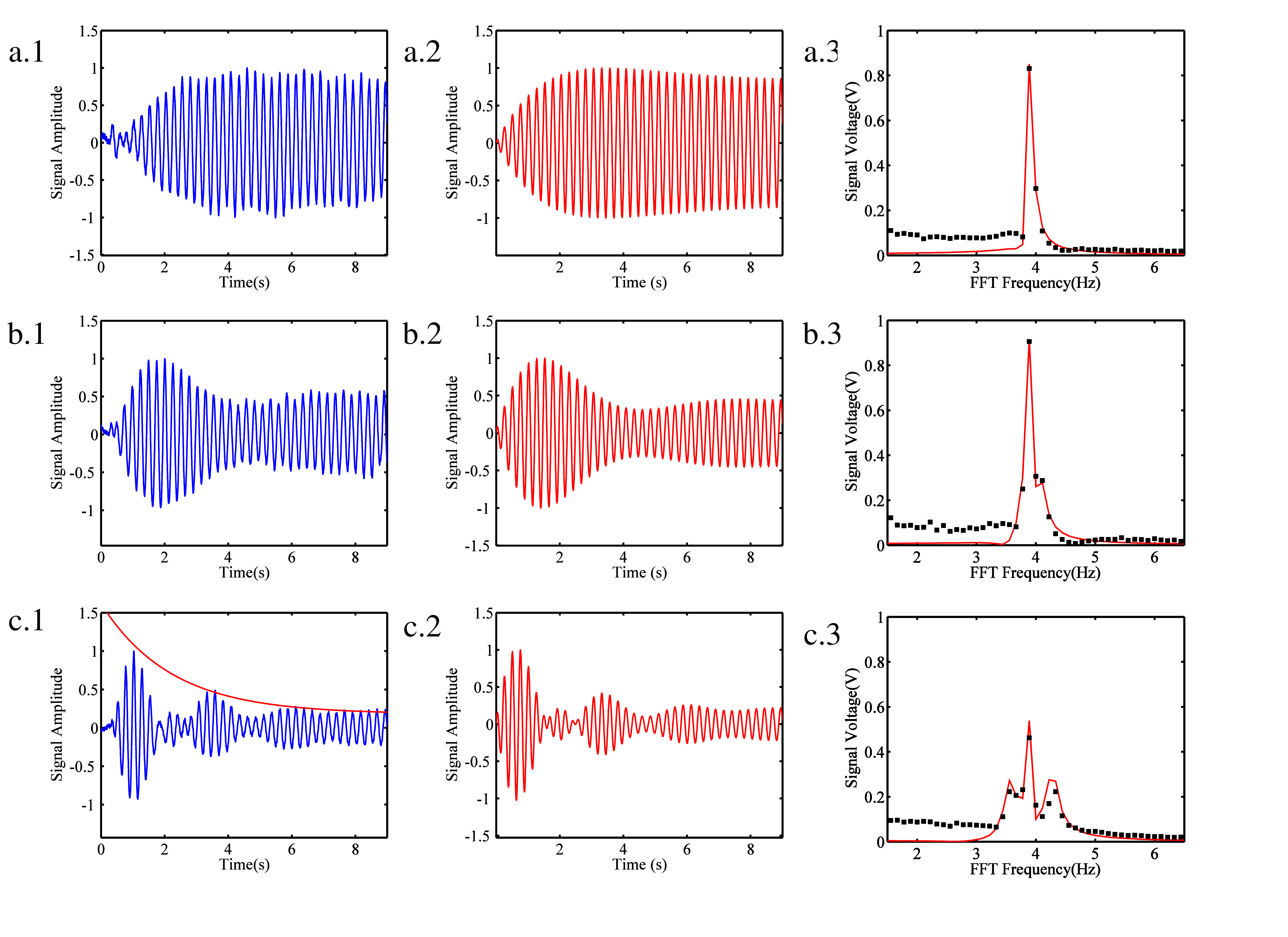}% Here is how to import EPS art
	\caption{\label{fig:TFdomain} Transient time-domain signals and the corresponding frequency-domain spectra when changing the oscillating magnetic field amplitudes. The static magnetic field $B_0=125$~nT and the oscillating magnetic field $B_{\rm M}=$ 3~nT (a), 9~nT (b) and 21~nT (c), respectively. The blue lines and black squares in the figures are from experimental data, while the red lines are from the simulated curves.  
	}
\end{figure*}

The pump and probe beams continuously interact with the atoms, and the frequency of $B_{\rm M}$ is set as $\omega =\omega_{\rm g}$. Here, the oscillating magnetic field is pulsed and with a duration of 10 seconds. The measured signal with Larmor frequency is shown in Fig.~\ref{fig:TFdomain}, both in the time and frequency domain. The simulation (red line) and experimental results (blue line and black square) are shown in Figs.~\ref{fig:TFdomain}~(a.1-a.3) and obtained when $\Omega_{\rm R} \approx \Gamma_{\rm g}$. Only one peak of Larmor frequency signal appears in the frequency domain, and no MT signal exists. The data for $\Omega_{\rm R} \approx 2\Gamma_{\rm g}$ in Figs.~\ref{fig:TFdomain}~(b.1-b.3) show that the time-domain signal envelope appears and the frequency-domain signal remains one peak, which shows both the dressed-spin effect in the time domain while the MT is not observable in the frequency domain. In the case of $\Omega_{\rm R} \approx 4\Gamma_{\rm g}$ shown in Figs.~\ref{fig:TFdomain}~(c.1-c.3), the dressed-spin effect in the time domain and the MT signal in the frequency domain are apparent. To conclude, although the MT is a kind of dressed-spin effect in the frequency domain, the distinguished MT signal requires the Rabi frequency $\Omega_{\rm R} \gg\Gamma_{\rm g}$, while the dressed-spin effect in the time domain can be observed when $\Omega_{\textrm{R}}$ is comparable with $\Gamma_{\rm g}$. In the experiments, the characteristic evolution time of the transverse angular momentum or the coherence time of the states is measured to be 2~s, see Fig.~\ref{fig:TFdomain}~(c.1).

\begin{figure*}
	\includegraphics[width=0.9\textwidth]{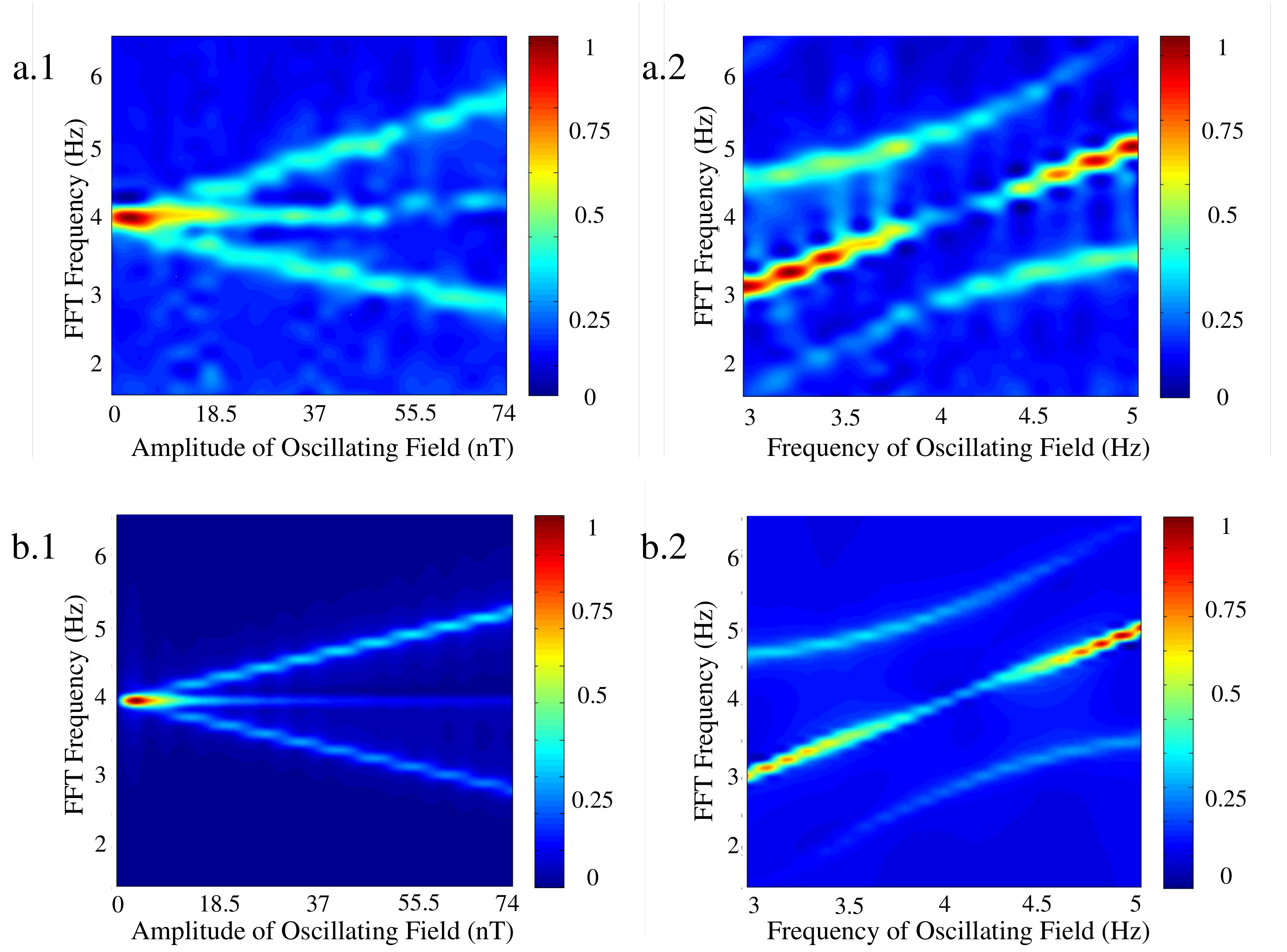}% Here is how to import EPS art
	\caption{\label{fig:MTC8C9} Dependence of FFT spectrum on changing the amplitude and frequency of the oscillating magnetic field for experiment (a) and simulation (b). The amplitude of oscillating magnetic field is about 70~nT  for Figs.~(a.2) and (b.2), and the frequency of the oscillating magnetic field is about 4~Hz for Figs.~(a.1) and (b.1).
	}
\end{figure*}

By changing both the amplitude and frequency of the oscillating magnetic field, the ULF-MT signals are shown in Fig.~\ref{fig:MTC8C9}. $\Omega_{\rm R}$ approximately behaves linearly with the amplitude of the resonant oscillating magnetic field $B_{\rm M}$, which agrees with $\Omega_{\rm R} =(1/2)\gamma_{\textrm{g}}B_{\textrm{M}}$ and indicates a method of measuring the amplitude for the ULF oscillating magnetic field. As shown in Fig.~\ref{fig:MTC8C9}~(a.2), one of the MT sidebands disappears in the case of far detuning ($\left| \omega-\omega_{\rm g}\right| \gg 0$), which agrees with the simulation result shown in Fig.~\ref{fig:MTC8C9}~(b.2), and the interval $\Delta$ of the MT satisfies the $\Delta=\sqrt{(\delta \omega)^2+(\Omega_{\rm R})^2}$, where the detuning $\delta \omega = \omega_{\rm g}-\omega$. The ratio of the sideband and the center peak is also influenced by collisions \cite{quoc2013stochastic,quoc2016mollow}, and to quantitify the phenomenone in $^{3}\textrm{He}$ needs further researches by controlling the collision rates. We notice that the amplitudes of two MT sidebands are same in experiments, but the amplitude of higher frequency sideband is larger in simulation, which needs further investigations.

\begin{figure*}
	\includegraphics[width=0.9\textwidth]{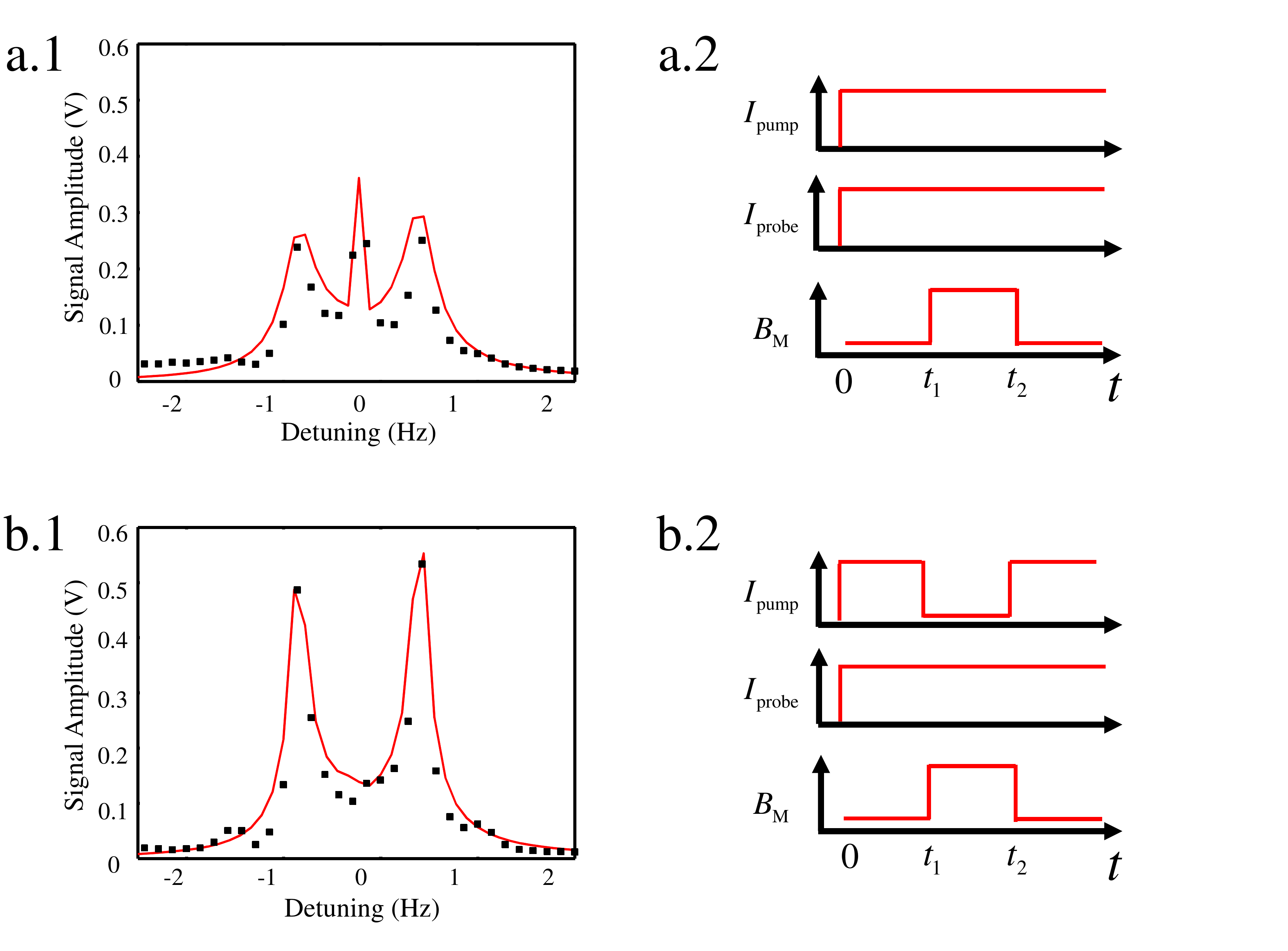}% Here is how to import EPS art
	\caption{\label{fig:onoffpump} Comparison of the continuously optical pumping with impulsively optical pumping. Figures.~\ref{fig:onoffpump}(a.1 and b.1) show the frequency-domain signals, and Figs.~\ref{fig:onoffpump}(a.2 and b.2) show the time-sequence diagrams of the pumping beam, the probe beam and the oscillating magnetic field. $t_{\textrm{2}}$ ($t_{\textrm{1}}$) is the start (stop) time of optical pumping beam, $t_{\textrm{1}}$ ($t_{\textrm{2}}$) is the start (stop) time of the oscillating magnetic field. The black squares (red lines) show the experimental (simulated) data. 
	}
\end{figure*}

In theory, the amplitude of MT center-peak signal is generated by the spontaneous emission \cite{barnett2002methods}. Figure~\ref{fig:onoffpump} shows the different MT signals with both continuously and impulsively running optical pumping, which is realized by using the shutter. The sequential time diagram is shown in Figs.~\ref{fig:onoffpump}(a.2 and b.2), where $t_{\textrm{2}}$ ($t_{\textrm{1}}$) is the start (stop) time of optical pumping beam, and $t_{\textrm{1}}$ ($t_{\textrm{2}}$) is the start (stop) time of the oscillating magnetic field. The center peak of MT disappears in Fig.~\ref{fig:onoffpump}(b.1), which indicates that the optical pumping is dominant on the creation of the center peak of MT in the experiments. The energy is transferred from the center peak to two sidebands of the MT by manipulating the optical pumping beam, while in the case where the spontaneous emission dominates the relaxation process, the manipulation is not obvious. \\

\section{\label{sec4}CONCLUSION}

We have observed the 4~Hz MT signal with 2~s coherent time, and push the frequency regime of MT observation by three orders of magnitude, from the radio
frequency to the ULF \cite{basler2015radio}. In order to measure the ULF signal, we utilize MECs between the ground and metastable states of $^{3}\textrm{He}$, which provides more efficient MT detection method. The ULF MT induced by the magnetic moment is described by the angular momentum equations, and the simulation results are in accordance with the experimental data. A new phenomenon in our experiments is that the center peak of the ULF MT can be controlled by the pump laser, which leads the possibility of realizing optical modulation. For example, we can utilize a pump laser, an oscillating magnetic field and an atomic cell to modulate the frequency of the probe laser, and control the sideband efficiency. Moreover, the frequency interval of the sidebands is linear with the amplitude of the resonant oscillating magnetic field, which satisfies $\hbar \Omega_{\rm R} = (1/2)\hbar\gamma_{\rm g}B_{\rm M}$, and indicates a possible method of measuring the amplitude of the ULF oscillating magnetic field.\\

\section*{\label{sec5}ACKNOWLEDGMENTS}

We thank for W.L., H.W. T.W., and H.d.W. with experimental and technical assistance. This project is supported by National Natural Science Foundation of China (61571018, 61531003, 91436210, 6157010573); National key research and development program.   

\nocite{*}

\bibliographystyle{unsrt}

\end{document}